\documentclass[twocolumn]{aastex702}
\usepackage{amsmath} 
\usepackage{graphicx}
\usepackage{hyperref}
\usepackage[toc,page]{appendix}
\usepackage{cleveref}

\defcitealias{Lim2013}{L13}

\begin{document}

\title{Hidden in the Halo: New Diffuse Star Clusters in the M82 Starburst with JWST}

\author[0000-0002-8217-5626]{Deepthi S. Prabhu}
\affiliation{Steward Observatory, University of Arizona, 933 North Cherry Avenue, Tucson, AZ 85721-0065, USA}
\email{dprabhu@arizona.edu}

\author[0000-0003-4102-380X]{David J. Sand}
\affiliation{Steward Observatory, University of Arizona, 933 North Cherry Avenue, Tucson, AZ 85721-0065, USA}
\email{dsand@arizona.edu}

\author[0000-0001-9649-4815]{Bur\c{c}in Mutlu-Pakdil}
\affil{Department of Physics and Astronomy, Dartmouth College, Hanover, NH 03755, USA}
\email{Burcin.Mutlu-Pakdil@dartmouth.edu}

\author[0000-0003-2599-7524]{Adam Smercina}
\affil{Department of Physics and Astronomy, Tufts University, 574 Boston Ave., Medford, MA 02155, USA}
\email{Adam.Smercina@tufts.edu}

\author[0000-0002-1763-4128]{Denija Crnojevi\'{c}}
\affil{Department of Physics and Astronomy, University of Tampa, 401 West Kennedy Boulevard, Tampa, FL 33606, USA}
\email{dcrnojevic@ut.edu}

\author[0000-0002-5434-4904]{Michael G. Jones}
\affiliation{IPAC, Mail Code 100-22, Caltech, 1200 E. California Blvd., Pasadena, CA 91125, USA}
\email{mgjones@ipac.caltech.edu}

\author[0000-0001-8354-7279]{Paul Bennet}
\affiliation{Space Telescope Science Institute, 3700 San Martin Drive, Baltimore, MD 21218, USA}
\email{pbennet@stsci.edu}

\author[0000-0001-9775-9029]{Amandine Doliva-Dolinsky}
\affiliation{Department of Physics, University of Surrey, Guildford GU2 7XH, UK}
\email{a.doliva-dolinsky@surrey.ac.uk}

\author[0000-0002-0956-7949]{Kristine Spekkens}
\affiliation{Department of Physics, Engineering Physics and Astronomy, Queen’s University, Kingston, ON K7L 3N6, Canada}
\email{kristine.spekkens@queensu.ca}

\author[0000-0003-4379-6777]{Debasish Hazarika}
\affil{Steward Observatory, University of Arizona, 933 North Cherry Avenue, Tucson, AZ 85721-0065, USA}
\email{debasish@arizona.edu}

\author[0000-0001-5368-3632]{Laura Congreve Hunter}
\affil{Department of Physics and Astronomy, Dartmouth College, Hanover, NH 03755, USA}
\email{Laura.C.Hunter@dartmouth.edu}

\author[0000-0002-4350-7632]{Jaclyn Jensen}
\affil{Department of Physics and Astronomy, Dartmouth College, Hanover, NH 03755, USA}
\email{Jaclyn.R.Jensen@dartmouth.edu}

\author[0009-0001-1147-6851]{Benjamin N.\ Velguth}
\altaffiliation{NSF Graduate Research Fellow}
\email{benjamin.n.velguth.gr@dartmouth.edu}
\affiliation{Department of Physics and Astronomy, Dartmouth College, Hanover, NH 03755, USA}

\author[0000-0002-5177-727X]{Dennis Zaritsky}
\affiliation{Steward Observatory, University of Arizona, 933 North Cherry Avenue, Tucson, AZ 85721-0065, USA}
\email{dennis.zaritsky@gmail.com}

\author[0009-0007-9488-7050]{Sasha N. Campana}
\affil{Department of Physics and Astronomy, Dartmouth College, Hanover, NH 03755, USA}
\email{Sasha.N.Campana.GR@dartmouth.edu}

\begin{abstract}

Diffuse star clusters (DSCs) occupy a poorly explored region of the size-luminosity plane between classical globular clusters (GCs) and dwarf galaxies. Here we report the discovery of four faint DSCs in the halo of the nearby starburst galaxy M82 using archival JWST/NIRCam imaging. We resolve individual stars in all four systems, whose color-magnitude diagrams are consistent with predominantly old ($\sim10$~Gyr) stellar populations, based on comparison with isochrones. DSC~1 hosts a  metal-poor population ($[\mathrm{M/H}]\sim-2$), while DSCs~2, 3 and 4 are better represented by $[\mathrm{M/H}]\sim-0.8$ isochrones. To establish their nature, we measure their sizes and integrated magnitudes using curve-of-growth and radial profile fits, which yield consistent results between these methods. The sizes of DSCs range from $\sim$~5.4 to 10.5~pc, making them more extended than typical GCs (2-3~pc). Their luminosities span M$_{\rm V}\simeq-4.1$ to $-6.4$, reaching $\sim$~1 mag fainter than the previously known M82 star cluster (SC) population. At these low luminosities, the systems occupy a sparsely sampled region of the size-luminosity plane where the distinction between SCs and dwarf galaxies becomes increasingly ambiguous. Most notably, DSC~3 is the faintest SC candidate identified in M82 to date and approaches the regime occupied by Milky Way ultra-faint compact satellites. These faint DSCs were identified through visual inspection alone, suggesting they are likely underrepresented in existing star cluster catalogs, and that a JWST-based archival search of nearby galaxies would be fruitful. 

\end{abstract}


\section{Introduction} 
\label{sec:intro}
Star clusters (SCs) are important components of galaxies that represent tracers of their star-formation, chemical-enrichment, and assembly histories (see \citealp{BrodieStrader2006}, \citealp{Forbes2018a}, \citealp{Adamo2020}, for detailed reviews). While globular clusters (GCs) are generally compact stellar systems with typical sizes of 2-3~pc \citep{Liu2016}, observations have revealed populations of unusually faint clusters with more extended structures \citep[e.g.,][]{Larsen2000,Peng2006,Howell2026}.

The high spatial resolution provided by Hubble Space Telescope (HST) ACS imaging was instrumental in identifying one class of such systems, called ``faint fuzzies'' (FFs), in the lenticular galaxy NGC~1023 \citep{Larsen2000,Brodie2002}. These are conventionally defined as extended, faint and red SCs with half-light radii (r$_{\rm h}$) greater than 7~pc, M$_{\rm V}>-$7, and $\rm (V-I)>1$ \citep{Larsen2000}. Although initially thought to be associated primarily with lenticular galaxies, similar systems, also commonly referred to as extended clusters (ECs), have since been identified in a wide variety of environments. These include the Milky Way \citep{vandenbergh2004}, M31 \citep{Huxor2005,Mackey2006}, M33 \citep{Stonkute2008,Cockroft2011}, the spiral galaxies M51 \citep{Chandar2004,Scheepmaker2007}, M81 \citep{Mayya2014}, M101 \citep{Simanton2015} and NGC~5195 \citep{Hwang2006}, and early-type galaxies such as Centaurus~A \citep{Mouhcine2010} and those in the Virgo and Fornax clusters \citep{Peng2006,Liu2016}. ECs are also found in much lower-mass galaxies, including NGC~247 \citep{Romanowsky2023}, the dwarf elliptical Scl-dE1 \citep{dacosta2009}, the dwarf irregular galaxies NGC~6822 \citep{Hwang2006,Howell2026} and IC~10 \citep{Howell2026}, the Fornax dwarf spheroidal galaxy \citep[dSph;][]{Pace2021, Crociati2026_fornax} and the M31 dSph, Andromeda~XXV \citep[Gep~I; ][]{Cusano2016, Crociati2026_gepI}. Remarkably, an EC has also been found to be associated with the ultra-faint Milky Way satellite Eridanus~II \citep{Crnojevic2016}. Together, these discoveries demonstrate that FFs/ECs are not restricted to a particular galaxy type.

As the known population has continued to expand, it has become clear that these diffuse systems span a broader range of sizes, luminosities, and colors than encompassed by the original FF definition \citep[e.g.,][]{Peng2006,Bruns2012}. To describe this broader population, \citet{Peng2006} introduced the term \emph{diffuse star clusters} (DSCs) to describe low-luminosity systems (M$_{\rm V}>-8$), spanning a wide range of sizes (r$_{\rm h}\sim3$--30~pc). As these classifications are based primarily on observed physical properties, their boundaries are not strictly defined, with EC and FF often used interchangeably in the literature and DSC encompassing a broader range of diffuse SCs beyond the specific criteria of the original FF definition. We therefore use DSC as the general term throughout this work, while retaining EC or FF when referring to systems described as such in previous studies. 

Beyond their structural diversity, DSCs also show considerable variation in their stellar populations and inferred origins. The HST/ACS study of early-type galaxies in the Virgo Cluster found predominantly red DSC populations, with colors consistent with ages $\gtrsim5$ Gyr, high metallicities, or a combination of the two \citep{Peng2006}. The ECs in NGC~1023 have likewise been found to be old ($\sim7$--$9$ Gyr) and metal-rich, with a mean $[\mathrm{Fe/H}]=-0.40\pm0.29$ \citep{Lopezsantamaria2025}. By contrast, spectroscopy of four ECs in NGC~6822 showed them to be similarly old ($\gtrsim8$ Gyr) but substantially more metal-poor ($[\mathrm{Fe/H}]\lesssim-1.5$), with their stellar populations and kinematics suggesting a possible accretion origin \citep{Hwang2014}. The EC associated with the ultra-faint dwarf Eridanus~II is also very metal-poor, with $[\mathrm{Fe/H}]\simeq-2$, as measured from the spectroscopy of its member stars and as inferred from isochrone matching \citep{Zoutendijk2020,Simon2021}. A different case is Fornax~6 in the Fornax dSph, which has been found to be relatively metal-rich, with $[\mathrm{Fe/H}]=-0.71\pm0.05$, but substantially younger, with an age of $\sim2$ Gyr inferred from comparison with stellar isochrones \citep{Pace2021}.

While these measurements provide increasingly detailed constraints on individual systems, the formation and evolution of DSCs remain uncertain. One proposed pathway is the merging of compact clusters within larger SC complexes, which can produce long-lived, extended stellar systems \citep{Fellhauer2002,Bruns2009,Bruns2011}. Alternatively, the preferential occurrence of DSCs in disks, low-mass galaxies, and galactic halos suggests that DSC formation may be favored in relatively low-density environments \citep{Elmegreen2008,Liu2016}. Galaxy interactions and mergers may also create conditions favorable for the formation of DSCs \citep{Burkert2005,Liu2016}. Another possibility is that DSCs develop through the long-term evolution of initially more compact clusters, as stellar mass loss and internal dynamical processes can drive substantial cluster expansion \citep{Gieles2010,Madrid2012,Gieles2021}. 

As a nearby starburst galaxy in the M81 group, M82 \citep[NGC 3034; M$_{\rm dyn}$~$\sim$~10$^{10}$~M$_{\odot}$; ][]{Greco2012} provides a particularly interesting environment in which to investigate DSCs, with its recent star- and cluster-formation history closely linked to interactions within the group \citep[e.g.,][]{Yun1994,Mayya2008,Lim2013}. The HST/ACS imaging of the M82 disk by \citet{Mayya2008} led to the identification of 653 SCs, demonstrating the richness of its disk cluster population. Despite this large population, comparatively few old halo GCs have been spectroscopically identified in M82. \citet{Saito2005} reported two GC candidates, while \citet{Konstantopoulos2009} identified two additional potential old GCs. Using HST imaging, \citet[hereafter \citetalias{Lim2013}]{Lim2013} subsequently identified more than 1000 SCs, dominated by an intermediate-age disk population, together with 35 halo cluster candidates whose colors are consistent with predominantly old stellar populations. Their cluster size distribution peaks at an effective radius of $\sim$~2~pc but does extend beyond 10~pc, while several of the halo candidates were partially resolved into individual stars with HST/ACS. More recent wide-field searches combining archival ground and space-based data have identified additional GC candidates at large projected distances ($>$ 5~kpc) from M82 \citep{Pan2022}, further indicating that its outer cluster population remains incompletely characterized.

The sensitivity and angular resolution of the James Webb Space Telescope (JWST) provide an excellent opportunity to extend the census of M82 SCs toward fainter and more diffuse systems, while resolving their constituent stars. In this work, we report the discovery of four new DSC candidates in the field of M82 using deep archival JWST NIRCam imaging survey data. We investigate their resolved stellar populations, structural properties, and integrated luminosities, and place them in the context of SCs in M82 and other nearby galaxies. Throughout this work, we adopt a distance of 3.55~Mpc to M82 \citepalias[$\rm (m-M)_0=27.75\pm0.07$~mag;][]{Lim2013}\footnote{At this adopted distance of M82, $1\arcsec$ corresponds to 17.2~pc.}.

The remainder of this paper is organized as follows. Section~\ref{sec:data} describes the JWST observations and data reduction. Section~\ref{sec:discovery} describes the discovery and identification of the new DSC candidates. Section~\ref{sec:phot_cmd} presents resolved star photometry and color-magnitude diagrams (CMDs), while Sections~\ref{sec:size} and \ref{sec:int_phot} describe the estimation of sizes and integrated photometry for these systems. We present and discuss our results in Section~\ref{sec:results}, and summarize our main findings and conclusions in Section~\ref{sec:conclusions}.

\section{Observations}
\label{sec:data}

We make use of archival JWST observations of the M82 field obtained as part of the \textit{Cibola} Survey (GO 5145; PI: A. Smercina). The global stellar photometry and initial results for \textit{Cibola} will be presented in A. Smercina et al. (in prep). The NIRCam \citep{nircam} imaging mosaic covers a field of approximately $13.6'\times6.7'$ centered on M82, extending to projected galactocentric distances of $\sim8$~kpc. The data were obtained between 2025 March 15 and 22 in four filters: F115W, F200W, F335M, and F444W. In this work, we use only the short-wavelength F115W and F200W imaging, which provides the spatial resolution and sensitivity required to identify faint stellar systems and resolve their brightest constituent stars in the M82 field. The specific datasets that we used can be found in the Mikulski Archive for Space Telescopes (MAST): \dataset[https://doi.org/10.17909/4dyv-0432]{https://doi.org/10.17909/4dyv-0432}.

The NIRCam observations were obtained using a four-position \texttt{INTRAMODULEBOX} dither pattern. The individual exposures have effective integration times of 1513.9~s in F115W and 1352.8~s in F200W, corresponding to nominal exposure times per pointing of 6055.5~s and 5411.3~s, respectively. We use the Level~3 (\texttt{i2d}) mosaics for the initial identification of the cluster candidates (see Section~\ref{sec:discovery}) and for the integrated-light measurements described in Sections~\ref{sec:size} and \ref{sec:int_phot}. For the resolved-star photometry, we use the individual calibrated Stage~2 (\texttt{cal}) exposures, as described in Section~\ref{sec:phot_cmd}.

\section{Discovery of the New DSC Candidates}
\label{sec:discovery}

The cluster candidates were identified through visual inspection of the F115W and F200W mosaics. The search was carried out with the goal of identifying faint, diffuse stellar systems that are partially resolved into individual stars while also exhibiting an underlying unresolved stellar component. The \texttt{i2d} mosaic images were examined over a range of spatial scales and contrast levels to identify semi-resolved systems that might be overlooked using a single display stretch. During this search, sources associated with previously identified SCs in the HST-based catalog of \citetalias{Lim2013} were excluded.

\begin{figure*}[t]

\centering
\includegraphics[width=\linewidth]{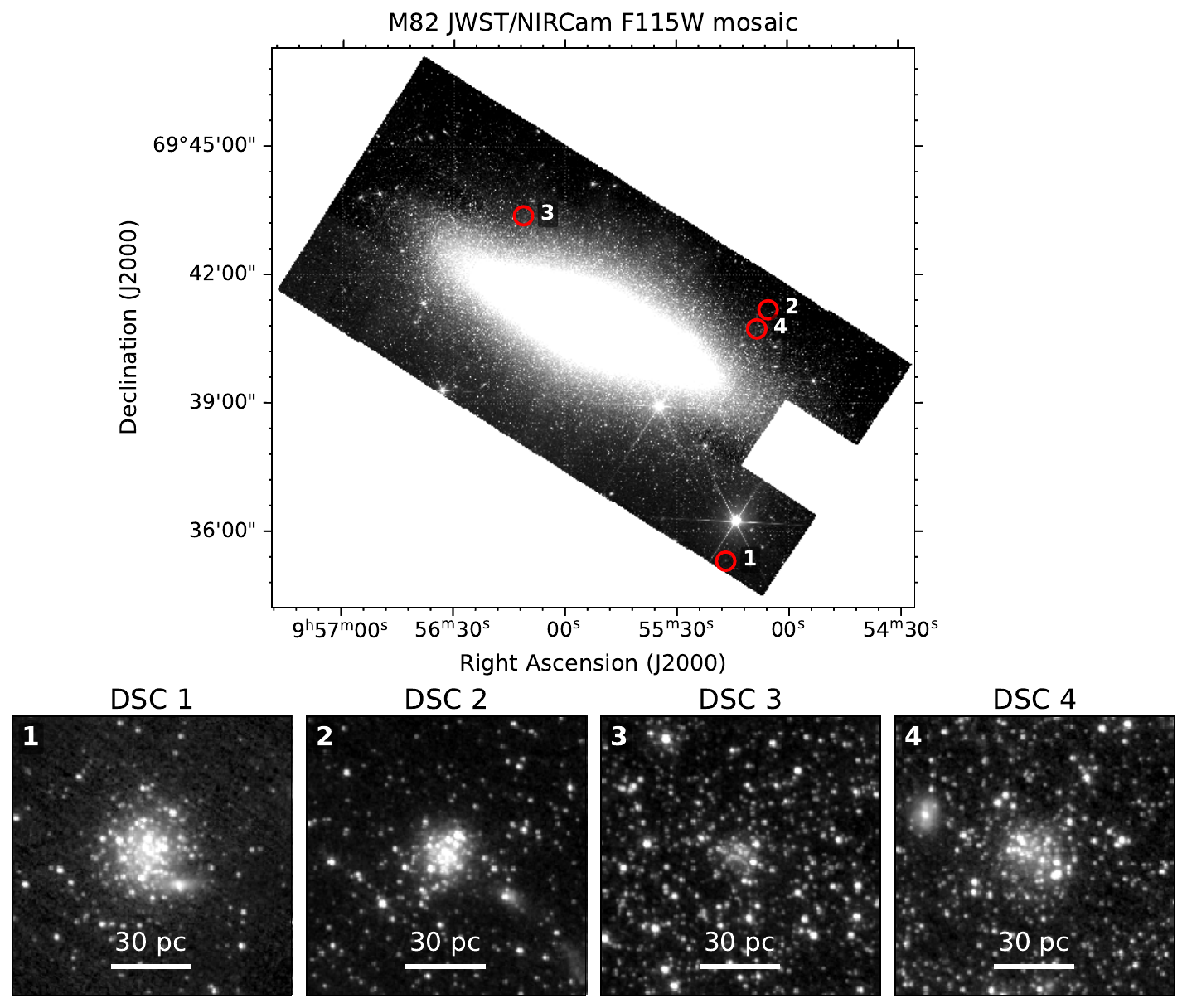}
\caption{Top: The F115W mosaic image showing the M82 JWST NIRCam survey field. The four newly discovered DSC candidates are marked using red circles. Bottom: The zoomed-in cutout images ($6\arcsec \times 6\arcsec$) centered on each object.}
\label{fig:spatial}
\end{figure*}

We identified four objects that satisfied these selection criteria and are not present in the \citetalias{Lim2013} catalog. Their spatial locations on the F115W mosaic, together with zoomed-in cutout images centered on each object, are shown in Figure~\ref{fig:spatial}, while their coordinates and projected galactocentric distances (D$_{\rm {proj}}$) are listed in Table~\ref{tab:positions}. To place their locations in the context of the M82 SC population, we use the XY coordinate system centered on the galaxy, defined by \citetalias{Lim2013}, where X and Y represent the projected distances along the major and minor axes of M82, respectively. All four candidates have $|\rm Y|>0.88$~kpc and therefore fall within the halo region defined by \citetalias{Lim2013} (see Figure~\ref{fig:xy_appendix} in Appendix~\ref{appendix1}). Three of the candidates (DSC~1, 2, and 4) are clearly semi-resolved, displaying overdensities of resolved stars superposed on diffuse integrated light. DSC~3 is extremely faint and is only barely resolved amid the high stellar density of the surrounding field. Nevertheless, it shows a localized stellar overdensity associated with a diffuse component, similar to the other candidates.

Following their identification, we searched additional catalogs to determine whether the four candidates had been reported previously. We compared their positions with the M82 GC candidate catalogs from \citet{Saito2005}, \citet{Davidge2008}, \citet{Konstantopoulos2009}, \citet{Chies-Santos2022}, and \citet{Pan2022}, the catalog of SCs in the M82 disk presented by \citet{Li2015}, as well as the M81 group ultra-faint dwarf (UFD) galaxy candidate catalogs of \citet{Bell2022} and \citet{Gozman2024}. None of the four candidates is associated with a previously cataloged source in these studies.

\begin{table}
\centering
\caption{Positions and projected galactocentric distance of the M82 DSC candidates.}
\label{tab:positions}
\begin{tabular}{lccc}
\hline
\hline
Object & R.A. (J2000) & Decl. (J2000) & D$_{\rm {proj}}$ \\
      & (deg.)         & (deg.)       & (kpc) \\
\hline
DSC~1 & 148.820563 & 69.588599 & 6.48\\
DSC~2 & 148.772369 & 69.686463 & 4.24\\
DSC~3 & 149.046988 & 69.723150 & 3.18\\
DSC~4 & 148.785167 & 69.679035 & 3.94\\
\hline
\end{tabular}
\end{table}


\section{Resolved Stellar Populations} \label{sec:phot_cmd}

\subsection{Point-source Photometry and Completeness} \label{sec:photometry}

Point-spread function (PSF) photometry was performed using the \texttt{DOLPHOT} package \citep{dolphot} with the JWST/NIRCam module \citep{Weisz24}. Rather than performing photometry on the full mosaic image, we carried out independent reductions for each cluster candidate. For each object, we first identified all calibrated Stage~2 (\texttt{cal}) exposures whose sky footprints intersect a circular region of radius $10\arcsec$ centered on the candidate. These matched \texttt{cal} exposures were retrieved from MAST and used as the input images for the photometric analysis. Performing photometry on the individual calibrated exposures preserves the native PSF and avoids the correlated noise introduced during image resampling and drizzling. The \texttt{cal} exposures thus selected around each cluster cover a region approximately $2\arcmin\times2\arcmin$, encompassing both the cluster sources and surrounding control sources for background assessment. The corresponding Level~3 F115W (\texttt{i2d}) mosaic was adopted as the astrometric reference frame for each reduction.

Image preprocessing followed the procedures recommended in the \texttt{DOLPHOT} User Guide, and we adopted the parameter values optimized for JWST/NIRCam imaging by \citet{Weisz23,Weisz24}. In particular, we used \texttt{FitSky}~=~2 together with the recommended aperture, fitting, and sky annulus parameters. The resulting source catalogs were filtered using the standard \texttt{DOLPHOT} quality diagnostics to retain only well-measured point sources. Specifically, we required a signal-to-noise ratio of S/N~$\geq4$ in both filters, \texttt{ERROR FLAG}~$\leq2$, \texttt{OBJECT TYPE}~$\leq1$, $\mathrm{SHARP}^2 \leq 0.0225$, and crowding value, \texttt{CROWD}~$\leq0.5$ in each filter.
In addition to this, we used \texttt{Force1} = 1, which forces all detected sources to be classified as stars (either bright or faint, returning global object types 1 or 2). This setting improves photometric precision and detection completeness in extremely crowded fields.

The resulting photometry was corrected for foreground Galactic extinction as follows. Previous studies \citep{Dalcanton2009,Marion2015} have noted that the Galactic dust maps of \citet{Schlegel1998}, recalibrated by \citet{Schlafly2011}, are contaminated by emission from M82 itself, leading to an overestimate of the Milky Way foreground reddening toward the galaxy. We verified this by inspecting the dust maps, which show enhanced reddening coincident with M82. We therefore followed the approach adopted by \citet{Dalcanton2009} and \citet{Marion2015}, using a foreground reddening value of $E(B-V)=0.05$~mag derived from regions surrounding M82 that are free from contamination by the galaxy, and apply this uniform correction to the photometry of all four cluster fields. Since all of our DSC candidates are located in the M82 halo, where internal extinction is expected to be negligible \citepalias{Lim2013}, we do not apply any additional extinction correction associated with the galaxy.

We then quantified the photometric uncertainties and completeness using the built-in \texttt{DOLPHOT} artificial-star routines. Approximately $5\times10^{5}$ artificial stars were injected one at a time into each reduction and recovered using the identical photometric procedures and quality cuts applied to the science images. The resulting artificial-star catalogs were used to determine the photometric uncertainties and completeness for each cluster field. The 50\% and 90\% completeness limits for each DSC are listed in Table~\ref{tab:completeness}. We note here that the tabulated completeness estimates largely reflect the uncrowded regions around each DSC, while the increased crowding toward the DSC centers can limit the recovery of fainter sources.

\begin{table}
\centering
\caption{Field Completeness Limits for each DSC field}
\label{tab:completeness}
\begin{tabular}{ccccc}
\hline\hline
 & \multicolumn{2}{c}{50\% Completeness} &
   \multicolumn{2}{c}{90\% Completeness} \\
\cline{2-3} \cline{4-5}
DSC & F115W & F200W & F115W & F200W \\
 & (mag) & (mag) & (mag) & (mag) \\
\hline
DSC~1 & 28.34 & 27.02 & 26.71 & 25.29 \\
DSC~2 & 28.72 & 27.40 & 27.38 & 26.26 \\
DSC~3 & 28.30 & 27.08 & 26.94 & 25.73 \\
DSC~4 & 28.74 & 27.51 & 27.64 & 26.34 \\
\hline
\end{tabular}
\end{table}

\subsection{Color-Magnitude Diagrams} \label{sec:cmd}

\begin{figure*}[!ht]
    \centering
    \includegraphics[width=\columnwidth]{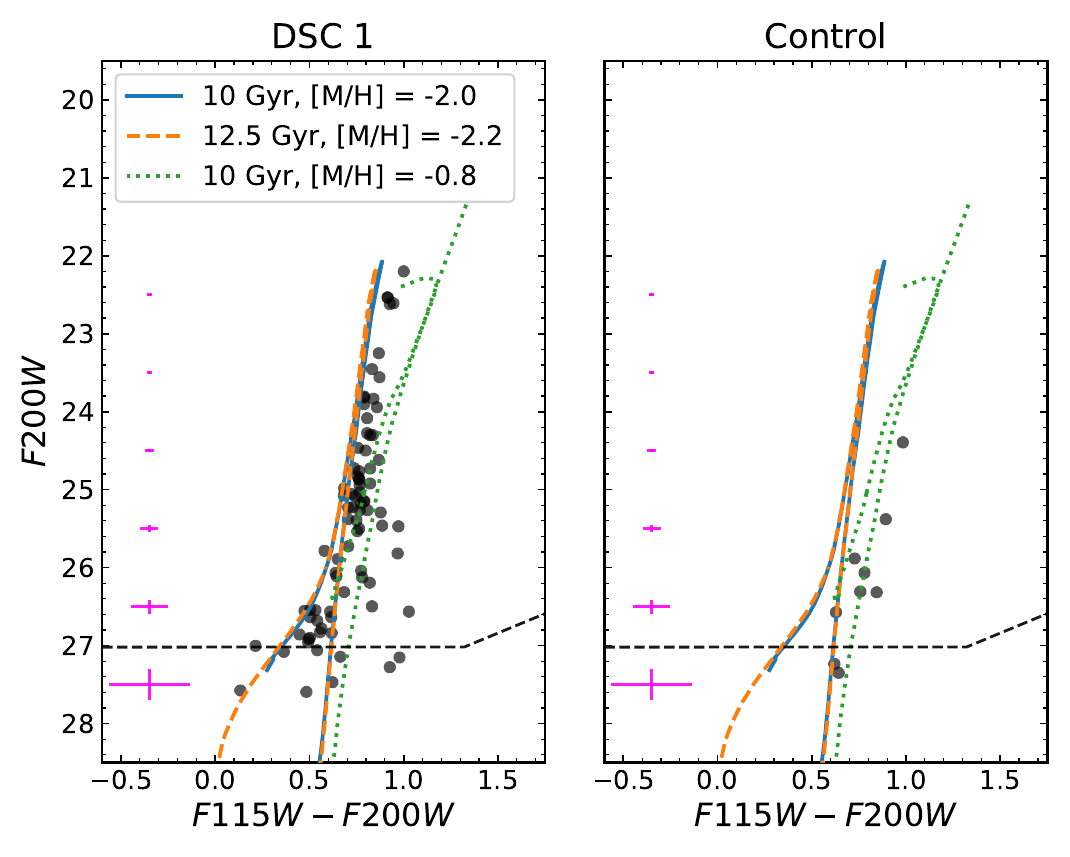} 
    \includegraphics[width=\columnwidth]{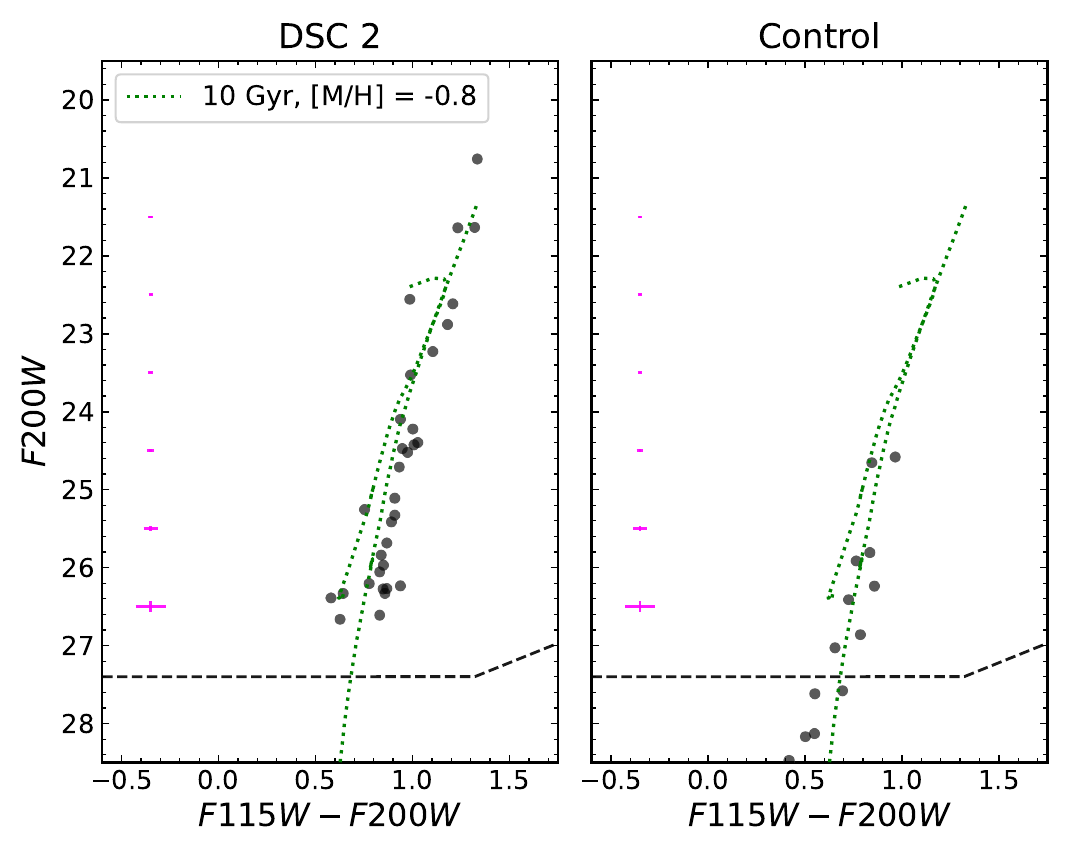} \\
    \includegraphics[width=\columnwidth]{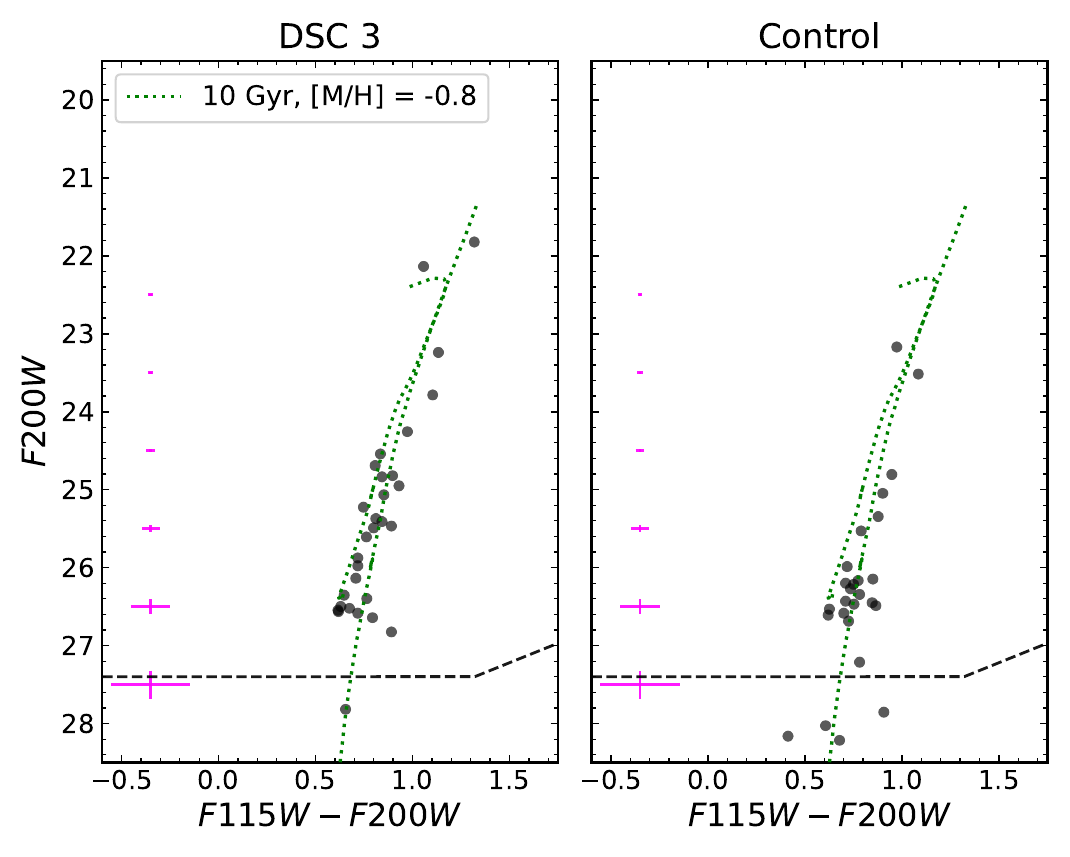} 
    \includegraphics[width=\columnwidth]{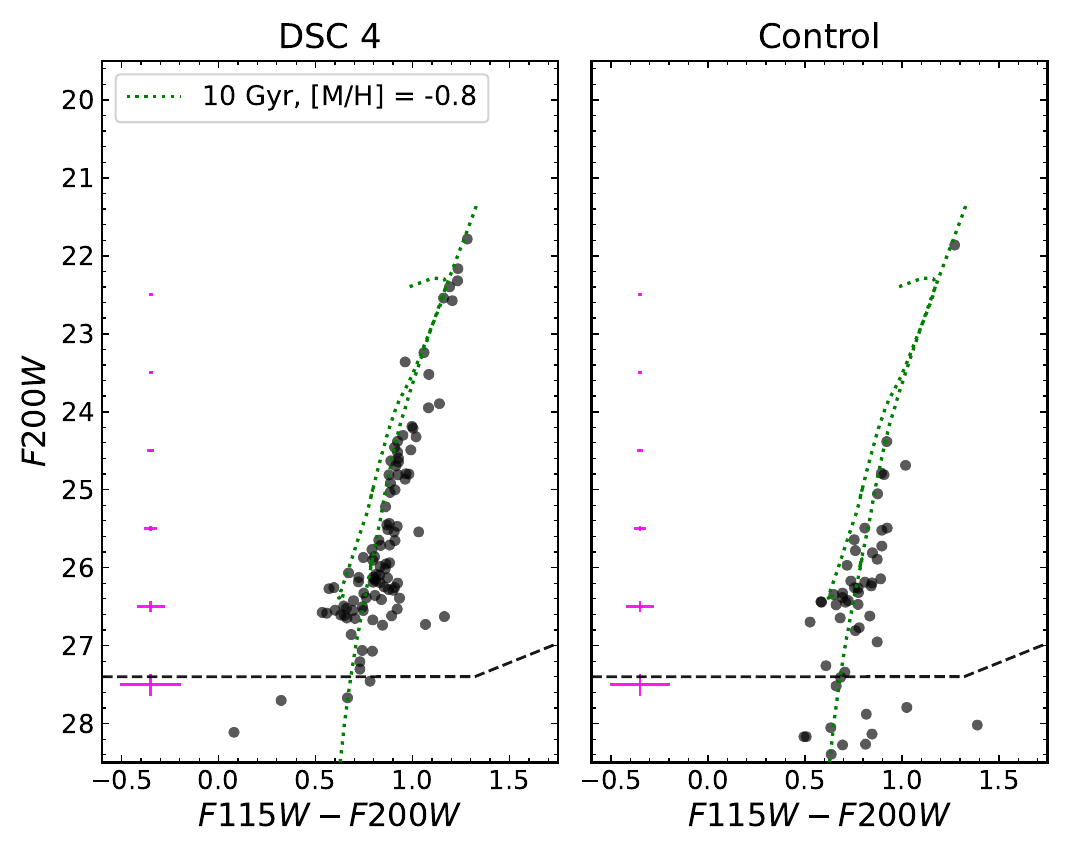} 
    \caption{CMDs of the newly discovered DSC candidates and the respective local equal-area control fields. The left panel of each figure set shows the CMD of stars within 2.5~r$_{\mathrm{h}}$ of each object (see Section~\ref{sec:size}), while the right panel shows the CMD of a representative equal-area control field at similar D$_{\rm {proj}}$ as each object. The black dashed line indicates the 50\%  completeness limit derived from ASTs for the field surrounding each object. However, we note here that increased crowding toward the DSC centers can limit the recovery of faint cluster stars before the nominal completeness limit is reached. The magenta error bars indicate the representative color and magnitude uncertainties derived from ASTs. PARSEC isochrones with the ages and metallicities reported in the legends, shifted to the adopted distance of M82, are overlaid as a reference for the stellar populations.}
    \label{fig:cmd}
\end{figure*}

The CMDs of the four cluster candidates and the respective local equal-area control fields representing the background are presented in Figure~\ref{fig:cmd}. For each object, the cluster CMD was constructed using all stars within a projected radius of 2.5~r$_{\mathrm{h}}$, where r$_{\mathrm{h}}$ is the final adopted value derived from the integrated-light analysis described in Section~\ref{sec:size}. The spatial distribution of the resolved sources included in each cluster CMD is shown in Appendix~\ref{appendix2}. To assess the level of field contamination, representative equal-area control fields were selected from each cluster catalog to match, as closely as possible, the projected galactocentric distance and local stellar density of each cluster while avoiding obvious stellar overdensities, diffuse sources, and artifacts.

PARSEC isochrones \citep{Bressan12, Chen2014, Tang2014, Chen2015, Marigo2017, Pastorelli2019} are overlaid on the CMDs as a guide to the underlying stellar populations. All the isochrones are placed at the adopted distance of M82 \citepalias[D=3.55~Mpc, ][]{Lim2013}. As noted in Section~\ref{sec:discovery}, all four DSC candidates lie within the halo region of M82. The SC population in this region was found to be predominantly consistent with old, metal-poor stellar populations, although a population extending toward redder colors was also identified \citepalias{Lim2013}. We therefore use old isochrones spanning metal-poor to intermediate metallicities as a physically motivated range for comparison with the observed CMDs. For DSC~1, we show old, metal-poor ($10$~Gyr, ${\rm [M/H]}=-2.0$ and $12.5$~Gyr, ${\rm [M/H]}=-2.2$) isochrones that are broadly consistent with the candidate cluster population, together with a more metal-rich (${\rm [M/H]}=-0.8$) isochrone representative of the surrounding M82 field population. For DSC~2, DSC~3, and DSC~4, a $10$~Gyr, ${\rm [M/H]}=-0.8$ isochrone provides a reasonable representation of the observed red giant branch (RGB). 

All four candidates show stars consistent with the RGB for an old stellar population at the adopted distance of M82, although the sequence is particularly sparse in DSC~3 and may be significantly affected by contamination from the surrounding M82 field. The photometry reaches the expected red clump (RC) level, with prominent concentrations of RC stars visible in DSC~1 and DSC~4. In DSC~2 and DSC~3, the RC is not clearly resolved as a distinct feature, likely owing to the greater crowding and field contamination in their surroundings, respectively. DSC~3 is the least resolved of the four systems, with only a small number of stars distinguishable above the dense underlying field.

Despite clear RGB detection in these systems, the number of resolved stars remains limited, particularly for the more crowded candidates. This precludes robust structural measurements of these systems based on star counts alone. We therefore measure their sizes and luminosities from integrated light, as described in the following sections.

\section{Cluster Sizes} 
\label{sec:size}

We employ two complementary approaches to measure the cluster sizes from their integrated
light. First, we use a non-parametric curve-of-growth (COG) method, similar to the technique used by \citet{Johnson2012}, to measure the r$_{\mathrm{h}}$. Second, we independently model the radial surface brightness profiles using a PSF-convolved exponential profile with a background component. Comparing the results from the two approaches provides an independent check on the derived r$_{\mathrm{h}}$ values.

\subsection{Curve-Of-Growth Analysis} \label{sec:cog}

\begin{figure*}[!ht]
\centering
\includegraphics[width=\linewidth]{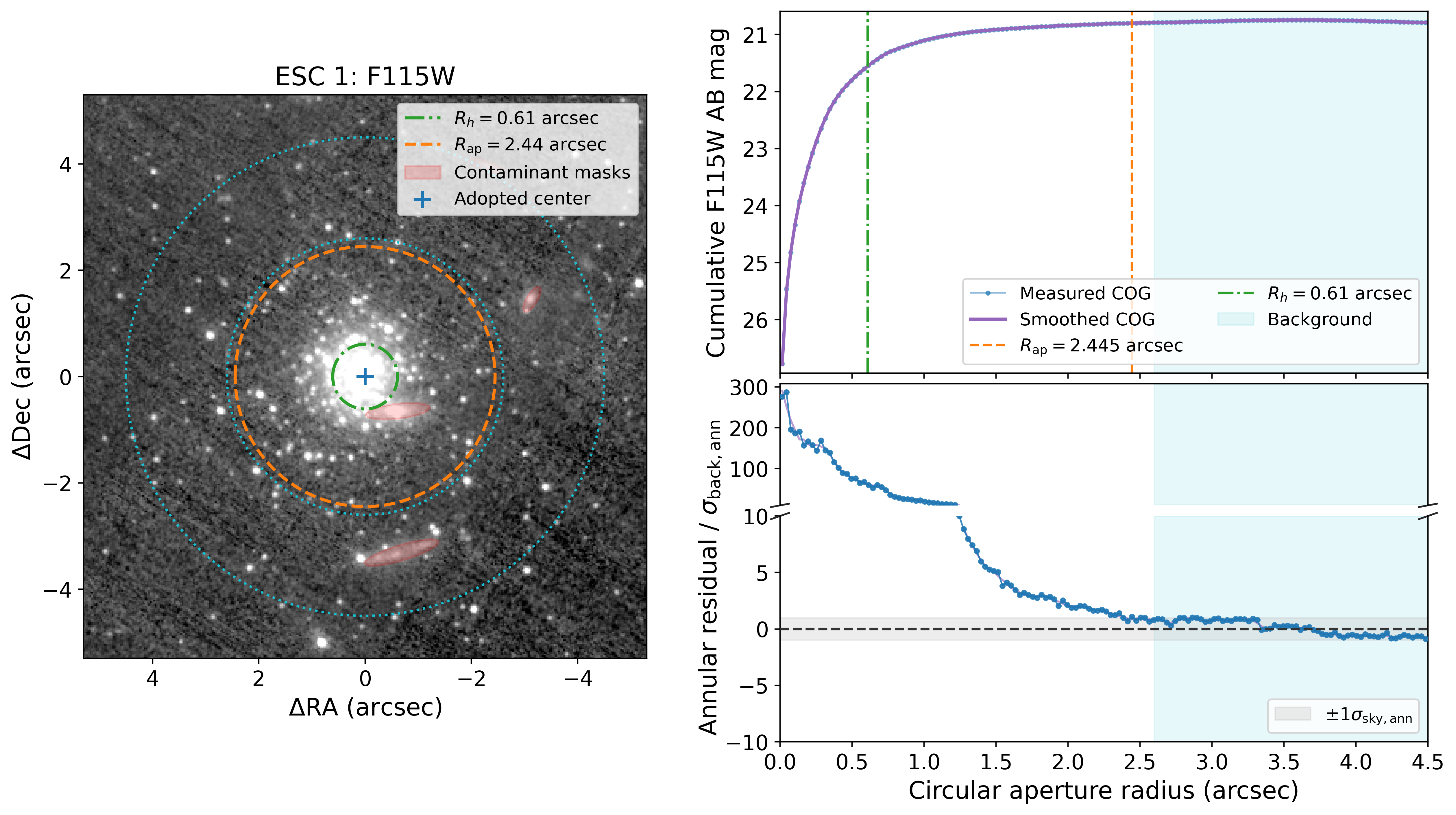}
\caption{Example of the curve-of-growth (COG) analysis for DSC~1 in the F115W image. \textit{Left:} Image cutout showing the adopted cluster center (blue cross), contaminant masks (red shaded ellipses), the measured r$_{\mathrm{h}}$ (green dash-dotted circle), the adopted photometric aperture R$_{\rm ap}$ (orange dashed circle), and the background annulus (cyan dotted circular annulus). \textit{Top right:} Background-subtracted cumulative F115W magnitude as a function of circular aperture radius. The measured and lightly smoothed COGs are shown for clarity. The adopted R$_{\rm ap}$ and r$_{\mathrm{h}}$ are indicated by the orange dashed and green dash-dotted vertical lines, respectively. \textit{Bottom right:} Background-subtracted annular residual, normalized by the dispersion of the background annuli ($\sigma_{\rm back, ann}$). The blue shaded region denotes the adopted background region used to estimate the local sky level.}
\label{fig:COG_esc1}
\end{figure*}

We begin by generating $20\arcsec\times20\arcsec$ image cutouts centered on each candidate and adopt the cluster center by visual inspection of the F115W
image. We construct COGs using circular aperture photometry independently in F115W and F200W filters, following approaches previously used for SCs \citep[e.g.,][]{Johnson2012,Howell2026}. Before performing the photometry, we mask obvious contaminating sources, including background galaxies and bright field stars. As an example, the adopted center and contaminant masks for the F115W cutout of DSC~1 are shown in the left panel of Figure~\ref{fig:COG_esc1}.

To construct the COG, we measure the flux in concentric annuli centered on the cluster with a radial width of $0.03\arcsec$. For annuli that are partially masked, the missing flux is estimated from the unmasked pixels, provided that at least 30\% of the annular area remains available. For annuli with less than 30\% usable area, the fluxes are interpolated from neighboring radial bins. From these measurements, we define an outer aperture radius, R$_{\rm ap}$, which represents the boundary of the cluster. Since the DSCs differ in their apparent size and surrounding field, we do not adopt a fixed aperture radius. Instead, we determine R$_{\rm ap}$ separately for each cluster and filter, allowing it to vary between $0.5\arcsec$ and $2.5\arcsec$ to encompass the apparent extent of the cluster while limiting contamination from the surrounding field. The specific criterion used to determine R$_{\rm ap}$ is described below.

We then measure the local background over $2.6\arcsec<r<4.5\arcsec$, immediately outside the range used to determine R$_{\rm ap}$. We divide this region into $0.1\arcsec$ annuli and measure the sigma-clipped mean intensity in each. We reject the annuli whose mean intensities deviate by more than $3\sigma$ from the median background level, and an area-weighted average of the remaining measurements is adopted as the local background. After background subtraction, the cumulative flux is converted to AB magnitude to construct the COG as a function of radius (see upper right panel of Fig.~\ref{fig:COG_esc1}).

To determine where the cluster signal reaches the background, we also calculate the annular residual, defined as the background-subtracted mean intensity in each annulus normalized by the dispersion of the annular background signal, $\sigma_{\rm back,ann}$. We adopt R$_{\rm ap}$ as the first minimum in the annular residual that is consistent with the background and coincides with a locally flat COG, similar to the empirical approach of \citet{Howell2026}.\footnote{Note that we smoothed the annular residual to be able to identify R$_{\rm ap}$. However, all photometric measurements are derived from the unsmoothed data.} The F115W COG and annular residual for DSC~1 are shown in the top- and bottom-right panels of Figure~\ref{fig:COG_esc1}, respectively. 

We then define r$_{\rm h}$, as the radius enclosing half of the background-subtracted flux within R$_{\rm ap}$ and determine it by interpolation of the cumulative-flux profile. The measured r$_{\rm h}$ is indicated by the green dash-dotted circle and vertical line in Figure~\ref{fig:COG_esc1}. This method provides a non-parametric estimate of the cluster size.

We estimate the uncertainty in the r$_{\rm h}$ measurement of each filter by accounting for contributions from the local background and adopted cluster center. To quantify the background contribution, we generate 1000 bootstrap resamples of the sigma-clipped mean background intensities and repeat the full COG analysis for each realization, redetermining both R$_{\rm ap}$ and r$_{\rm h}$. We use the 16$^{\rm th}$ and 84$^{\rm th}$ percentiles of the resulting r$_{\rm h}$ distribution to define the uncertainty interval. To account for uncertainty in the cluster center, we repeat the analysis after shifting the adopted center by five pixels in the cardinal and diagonal directions. The uncertainties associated with the background and cluster center are then combined in quadrature.

Finally, we perform the COG analysis independently in F115W and F200W and adopt the mean r$_{\rm h}$ as the final measurement. The final uncertainty combines the propagated errors from both filters with half the difference between their r$_{\rm h}$ measurements, added in quadrature. The resulting r$_{\rm h}$ values are converted to physical sizes assuming the adopted distance to M82 and are listed in Table~\ref{tab:esc_properties}.

\subsection{Radial Profile Fitting}

\label{sec:radial_profile}

As an independent estimate of the cluster sizes, we also fit their radial
surface brightness profiles. Unlike the COG analysis, this approach explicitly
accounts for the JWST/NIRCam PSF and fits the local background simultaneously
with the cluster profile.

We first construct empirical PSFs in F115W and F200W images by combining bright, isolated stars selected from the DOLPHOT photometric catalogs in the vicinity of the DSCs. For each DSC, we measure the radial profile in concentric circular annuli
of size 0.08$''$ centered on the adopted cluster position. The same contaminant masks used for the COG analysis are applied, and a sigma-clipped mean intensity is measured
in each annulus. The profiles are measured out to $4.5\arcsec$, with the
inner $3.0\arcsec$ used for fitting.

\begin{figure}[!htbp]
    \centering
\includegraphics[width=\linewidth]{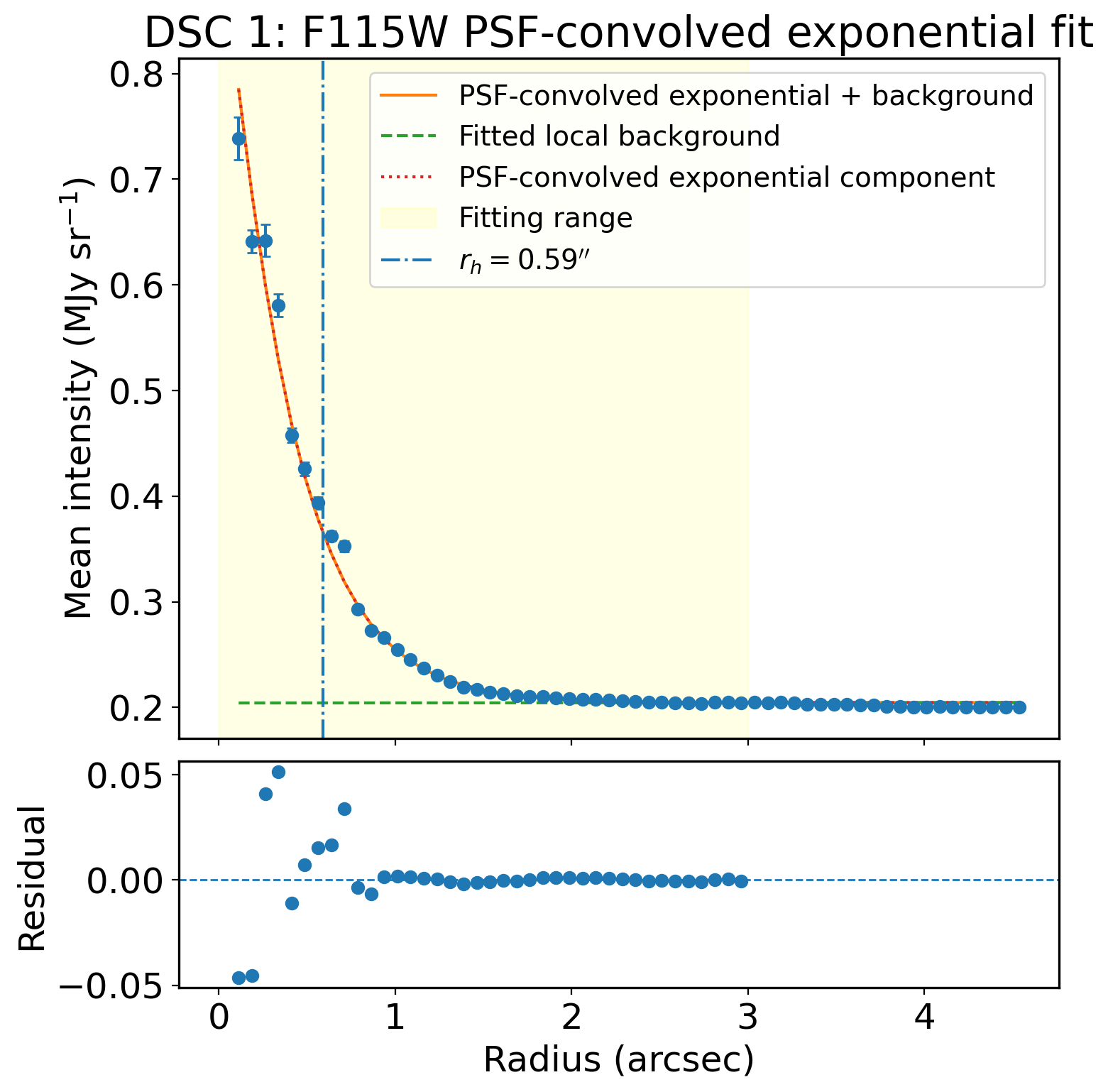}
\caption{Radial profile fit for DSC~1 using the F115W cutout. The blue points show the measured mean intensity in each annulus, and the orange curve shows the best-fitting PSF-convolved exponential model including the fitted local background. The shaded region indicates the radial range used for the fit, and the vertical dash-dotted line marks the
derived r$_{\rm h}$. The lower panel shows the residuals between the measured and best-fitting profiles.}
\label{fig:rad_profile}
\end{figure}

We fit the measured radial profiles with an exponential model of the form
\begin{equation}
    I(r) = I_0 \exp(-r/r_s),
\end{equation}
where I$_0$ is the central cluster intensity and r$_{\rm s}$ is the exponential scale radius. The exponential model is convolved with the empirical PSF before comparison with the observed profile. We also include a local background component in the fit, consisting of a constant level and a linear radial gradient. The model fitted to the radial profile can therefore be written as
\begin{equation}
    I_{\rm model}(r) =
    \left[{\rm PSF}\ast I(r)\right] + B_0 + B_1r,
\end{equation}
where $B_0$ and $B_1$ describe the local background. The radial profile of the
PSF-convolved model is measured using the same annuli and contaminant masks as for the data. We finally convert the fitted r$_{\rm s}$ to the r$_{\rm h}$ using the relation, r$_{\rm h}$ = 1.678~r$_{\rm s}$. As an example, we illustrate the radial profile fit for the F115W cutout of DSC~1 in Figure~\ref{fig:rad_profile}.

We determine the uncertainty on r$_{\rm h}$ from the profile fit and from
variations in the adopted center, fitting range, and contaminant masks. The
uncertainty from the profile fit is determined from the range of r$_{\rm s}$ values
consistent with the observed profile. We then repeat the fit after shifting the cluster center by five pixels in eight directions, changing the outer fitting radius from $3.0\arcsec$ to $2.5\arcsec$ and $3.5\arcsec$, and scaling the sizes of contaminant masks by factors of 0.9 and 1.1. For each case, we calculate the differences in r$_{\rm h}$
from the original fit and use their root-mean-square as the corresponding
uncertainty. These contributions are combined in quadrature with the uncertainty from the
profile fit. The resulting measurements are listed in Table~\ref{tab:esc_properties}.

\begin{table*}
\centering
\caption{Structural and photometric properties of the newly discovered DSCs.}
\label{tab:esc_properties}

\begin{tabular}{lcccc}
\hline\hline
Property & DSC~1 & DSC~2 & DSC~3 & DSC~4 \\
\hline
\multicolumn{5}{c}{COG Measurements} \\
\hline 
r$_{\mathrm{h}}$ (arcsec)              & 0.62$^{+0.02}_{-0.01}$ & 0.35$^{+0.05}_{-0.01}$ & 0.33$^{+0.05}_{-0.01}$ & 0.50$^{+0.02}_{-0.02}$\\
r$_{\mathrm{h}}$ (pc)                  & 10.70$^{+0.39}_{-0.23}$ & 6.00$^{+0.84}_{-0.19}$ & 5.75$^{+0.78}_{-0.23}$ & 8.56$^{+0.43}_{-0.35}$\\
F115W (AB mag)  & 20.71$^{+0.03}_{-0.04}$ & 21.33$^{+0.07}_{-0.02}$ & 22.44$^{+0.10}_{-0.21}$ & 22.07$^{+0.07}_{-0.09}$\\
F200W (AB mag)  & 20.88$^{+0.04}_{-0.04}$ & 21.36$^{+0.13}_{-0.05}$ & 22.50$^{+0.17}_{-0.28}$ & 22.04$^{+0.09}_{-0.07}$\\
M$_{\rm V}$ (Vega mag)                 & $-$6.48$^{+0.09}_{-0.09}$ & $-$5.53$^{+0.10}_{-0.06}$ & $-$4.40$^{+0.13}_{-0.24}$ & $-$4.81$^{+0.12}_{-0.13}$ \\

\hline
\multicolumn{5}{c}{PSF-convolved Exponential Profile Fit Measurements} \\
\hline
r$_{\mathrm{h}}$ (arcsec)              & 0.59$^{+0.04}_{-0.04}$ & 0.33$^{+0.04}_{-0.04}$ & 0.29$^{+0.03}_{-0.03}$ & 0.63$^{+0.04}_{-0.04}$\\
r$_{\mathrm{h}}$ (pc)                  & 10.24$^{+0.65}_{-0.65}$ & 5.60$^{+0.72}_{-0.72}$ & 5.00$^{+0.45}_{-0.45}$ & 10.83$^{+0.65}_{-0.65}$\\
F115W (AB mag)  & 20.87$^{+0.09}_{-0.09}$ & 21.63$^{+0.14}_{-0.14}$ & 23.02$^{+0.10}_{-0.10}$ & 21.79$^{+0.07}_{-0.07}$\\
F200W (AB mag)  & 21.06$^{+0.13}_{-0.13}$ & 21.69$^{+0.13}_{-0.13}$ & 23.22$^{+0.16}_{-0.16}$ & 21.89$^{+0.14}_{-0.14}$\\
M$_{\rm V}$ (Vega mag)                 & $-$6.31$^{+0.08}_{-0.08}$ & $-$5.19$^{+0.10}_{-0.10}$ & $-$3.73$^{+0.10}_{-0.10}$ & $-$5.01$^{+0.08}_{-0.08}$\\

\hline
\multicolumn{5}{c}{Adopted Values} \\
\hline
r$_{\mathrm{h}}$ (arcsec)              & $0.61_{-0.02}^{+0.03}$ & $0.34_{-0.02}^{+0.03}$ & $0.31_{-0.03}^{+0.03}$ & $0.56_{-0.07}^{+0.07}$ \\
r$_{\mathrm{h}}$ (pc)                  & $10.47_{-0.41}^{+0.44}$ & $5.80_{-0.42}^{+0.59}$ & $5.37_{-0.45}^{+0.59}$ & $9.70_{-1.19}^{+1.20}$ \\
F115W (AB mag)                & $20.79_{-0.09}^{+0.09}$ & $21.48_{-0.16}^{+0.17}$ & $22.73_{-0.31}^{+0.30}$ & $21.93_{-0.15}^{+0.15}$ \\
F200W (AB mag)                & $20.97_{-0.11}^{+0.11}$ & $21.52_{-0.18}^{+0.19}$ & $22.86_{-0.40}^{+0.38}$ & $21.96_{-0.11}^{+0.11}$ \\
M$_{\rm V}$ (Vega mag)              & $-6.39_{-0.10}^{+0.10}$ & $-5.36_{-0.18}^{+0.18}$ & $-4.07_{-0.36}^{+0.35}$ & $-4.91_{-0.12}^{+0.12}$ \\

\hline
\end{tabular}
\end{table*}

\section{Integrated-light photometry} \label{sec:int_phot}

We derive integrated F115W and F200W photometry independently using both the COG measurements and the PSF-convolved exponential profile fits described above. The two approaches estimate the total light in different ways: the COG method determines the asymptotic flux directly from the observed growth curve, whereas the profile-fitting method obtains the total flux by integrating the best-fitting PSF-convolved exponential model.

For the COG photometry, we measure the cluster flux within 3~r$_{\rm h}$,
where r$_{\rm h}$ is the mean of the F115W and F200W measurements, and adopt
the same radius in both filters. This choice includes most of the cluster
contribution while limiting contamination from the surrounding field. The local background is determined independently in each filter following Section~\ref{sec:cog}. We
then measure the background-subtracted flux within the aperture and use the COG to
correct for the flux outside it. The total flux is then taken as the median cumulative
value over the flat outer part of the COG. We then estimate an aperture correction as the ratio of this value to the flux within 3~r$_{\rm h}$. The resulting aperture corrections range from $-0.41$ to $+0.17$ mag, with a median of $-0.10$ mag. The largest corrections are found for the faintest cluster, DSC~3, which also has the lowest flux contrast compared to the surrounding field.

For the radial profile method, we instead obtain the total flux from the best-fitting PSF-convolved exponential model. Because the empirical PSFs are normalized to unit total flux, the convolution changes the shape of the model profile without changing its integrated flux. We therefore calculate the total F115W and F200W fluxes by integrating the intrinsic exponential component of the best-fitting model to infinite radius. 

The resulting total fluxes from both the methods are converted to apparent AB magnitudes and corrected for foreground MW extinction. We then convert the extinction-corrected apparent magnitudes to absolute magnitudes in F115W and F200W using the adopted distance to M82.

To facilitate comparison with previous studies of SCs, we also transformed the NIRCam photometry to the Johnson V band. Because this transformation depends on the underlying stellar population, we derived integrated $\rm V-\mathrm{F115W}$ and $\rm V-\mathrm{F200W}$ colors from PARSEC luminosity functions using ages and metallicities guided by the CMDs in Figure~\ref{fig:cmd}. Accordingly, for DSC~1 we use the two old, metal-poor populations shown in the CMD, while for DSCs~2, 3 and 4 we use the $10$~Gyr, $[\mathrm{M/H}]=-0.8$ population shown in the CMDs along with a slightly more metal-rich, $[\mathrm{M/H}]=-0.7$ model to account for uncertainty in the stellar-population choice. The extinction-corrected NIRCam AB magnitudes were first converted to the Vega system using the appropriate conversion factors \citep{Willmer2018}\footnote{\url{https://mips.as.arizona.edu/~cnaw/sun.html}}, and the model colors were then used to obtain independent estimates of M$_{\rm V}$ from F115W and F200W. For each photometric method, we adopt the mean of the M$_{\rm V}$ estimates inferred from the two NIRCam filters. We repeat this calculation for each adopted PARSEC model and use half the range of the resulting M$_{\rm V}$ values as the systematic uncertainty in the NIRCam to V transformation.

We estimate the photometric uncertainties separately for the two methods. For the COG measurements, we account for uncertainties associated with the local background, cluster centering, contaminant masking, and the choice of radial range used to measure the outer COG plateau. We also include the difference between the M$_{\rm V}$ estimates derived independently from F115W and F200W to account for filter-to-filter variations. For the radial-profile method, the flux uncertainty includes the uncertainties in the fitted model parameters, together with the effects of cluster centering, fitting range, and contaminant masking. For both methods, these contributions are combined in quadrature, and the final M$_{\rm V}$ uncertainties also include the systematic uncertainty associated with the adopted stellar-population models.

The F115W and F200W integrated magnitudes and the corresponding M$_{\rm V}$ estimates from both methods for all the DSC candidates are presented in Table~\ref{tab:esc_properties}.

\section{Results and Discussion} \label{sec:results}
\subsection{Stellar Populations}
\label{sec:stellar_pop}

The CMDs in Figure~\ref{fig:cmd} suggest that all four DSCs are dominated by old stellar populations, with some indication of a metallicity difference within the sample. DSC~1 is broadly consistent with old, metal-poor ($[\mathrm{M/H}]\sim-2$) PARSEC isochrones and appears more metal-poor than its surrounding field. DSCs~2 and 4 are instead better represented by the 10~Gyr, $[\mathrm{M/H}]=-0.8$ isochrone. The limited number of resolved stars in DSC~3 is also compatible with this isochrone, although its stellar population is less well constrained. More importantly, the CMDs of all four systems are consistent with the isochrones placed at the adopted distance of M82, providing strong evidence for their association with the galaxy.

These stellar properties are consistent with what is already known about the M82 halo cluster population. \citetalias{Lim2013} found a substantial population of old halo clusters, with the dominant population consistent with an age of $\sim$10~Gyr and a low metallicity of Z=0.0001 ([M/H] $\simeq$ $-$2.19). Their sample also includes clusters extending toward the colors predicted for more metal-rich stellar
populations. The apparent difference between DSC~1 and others may therefore reflect a metallicity range similar to that already seen among M82 halo clusters. 

A broad range of metallicities is also found among the diffuse and/or extended SCs in other galaxies. The ECs in NGC~6822 are old and metal-poor, with ages $\gtrsim8$~Gyr and [Fe/H] $\lesssim$ $-$1.5 \citep{Hwang2014}, whereas the old ECs in NGC~1023 are considerably more metal-rich, with a mean [Fe/H] $\simeq$ $-$0.4 \citep{Lopezsantamaria2025}. The populations indicated by our CMDs therefore fall within the range observed among DSCs in other environments. Given the small number of resolved stars and the age-metallicity degeneracy along the RGB, however, we treat the isochrone comparisons as qualitative and do not derive formal ages or metallicities.

\subsection{Nature of the Systems}
\label{sec:nature}

As seen in Table~\ref{tab:esc_properties}, the COG and PSF-convolved exponential profile fitting yield generally consistent sizes and integrated magnitudes for all four DSCs. For each quantity, we adopt the mean of the two measurements as the final value. The uncertainties from the individual measurements are propagated to the mean, and half the difference between the two methods is included in quadrature to account for method-dependent variation.

\begin{figure*}[t!]
    \centering
\includegraphics[width=0.8\textwidth, height=0.6\textwidth]{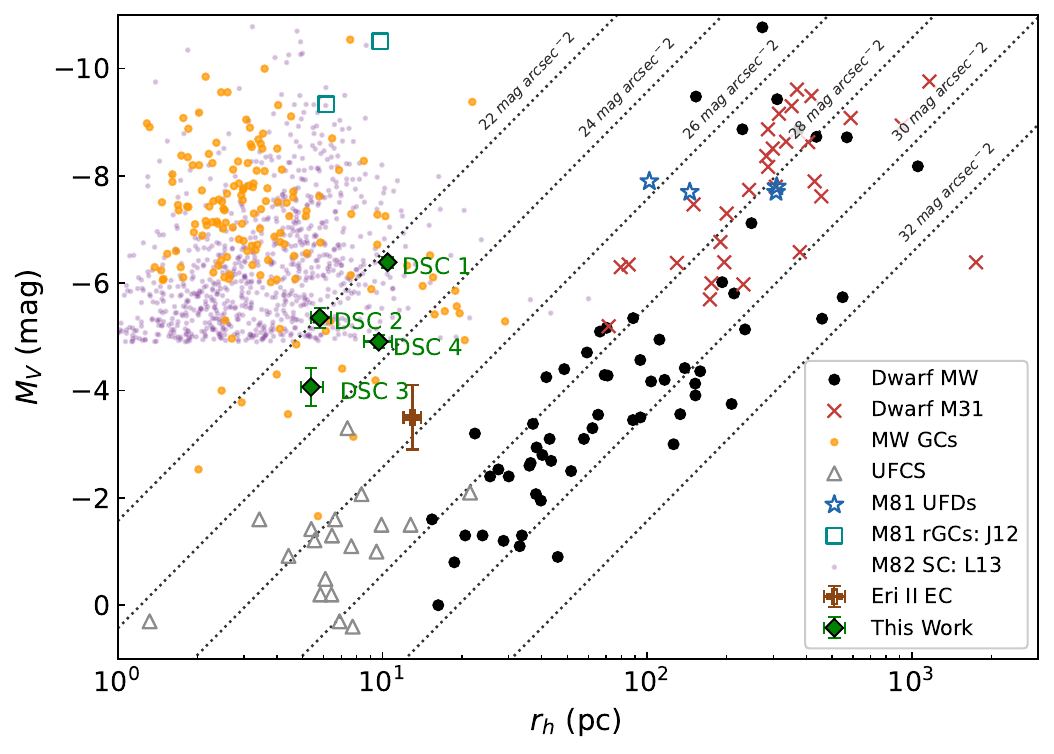}
\caption{Half-light radius (r$_{\rm h}$) versus absolute $V$-band magnitude (M$_{\rm V}$) for the four DSCs identified in this work (green diamonds) and comparison stellar systems. M82 SCs from \citetalias{Lim2013} and the remote M81 Group GCs from \citet{Jang2012} are shown as purple circles and light-blue open squares, respectively. Milky Way GCs and UFCSs are shown as orange open circles and gray triangles, respectively. Satellite dwarf galaxies of the Milky Way, M31, and the M81 Group are shown as black circles, brown crosses, and blue stars, respectively. The EC associated with the Eridanus~II UFD \citep{Crnojevic2016,Simon2021} is marked separately with a brown cross. The dotted lines indicate lines of constant surface brightness.}
\label{fig:size_lum}
\end{figure*}

Figure~\ref{fig:size_lum} places the four systems in the size-luminosity plane alongside SCs and faint stellar systems in the Local Volume. The Milky Way GCs, ultra-faint compact satellites (UFCSs), and Milky Way and M31 UFDs are drawn from the Local Volume Database \citep[LVDB v1.1.0;][]{Pace2025}, while the M81 UFDs are from \citet{Gozman2024}. We also include the M82 SC sample from \citetalias{Lim2013} and the two remote M81 Group GCs identified by \citet{Jang2012}. For the \citetalias{Lim2013} sample, we convert the reported apparent V magnitudes to M$_{\rm V}$ using the distance and foreground extinction adopted in this work, ensuring a consistent comparison with our systems.

It is evident from  Figure~\ref{fig:size_lum} that our candidates lie at lower luminosities and have larger sizes than the bulk of the Milky Way GC population. DSCs~1 and 4, with r$_{\rm h}\simeq10$~pc and M$_{\rm V}>-7$, satisfy the size and luminosity criteria commonly defined for FFs. DSCs~2 and 3 are more compact, with sizes between 5 to 6~pc, making them smaller than typical FFs but still larger than the typical r$_{\rm h}$ values of GCs. The size diversity within our sample therefore favors the broader DSC classification. 

Compared with the previously known M82 SC population, our clusters are notable for their low luminosities and extended sizes. The SCs identified by \citetalias{Lim2013} have a median r$_{\rm h}$ of $\sim3$~pc, whereas our sample spans r$_{\rm h}\simeq$~5 to 10~pc and reaches substantially fainter
luminosities. Previous spectroscopic and imaging studies have established the presence of old SCs from the inner regions of M82 to its extended halo \citep{Saito2005,Konstantopoulos2009,Lim2013,Pan2022}. In particular, \citetalias{Lim2013} identified 35 predominantly old SC candidates in the halo, including 13 systems in which a few individual stars could be resolved. We likewise resolve stars in our candidates, whose CMDs, as discussed in Section~\ref{sec:stellar_pop}, provide compelling evidence for their association with M82, with all four systems lying within the halo region defined by \citetalias{Lim2013}. Our systems therefore extend the known M82 halo SC population into a sparsely sampled region of the size-luminosity plane.

DSC~3 marks the most extreme case among our newly discovered cluster sample. Its low luminosity (M$_{\rm V}=-4.07^{+0.35}_{-0.36}$) makes it the faintest SC candidate identified in M82 to date, and its particularly small size (r$_{\rm h}\simeq5.4$~pc) places it close to the UFCS regime. This comparison is particularly relevant because the distinction between SCs and dwarf galaxies becomes increasingly ambiguous at such low luminosities. Recent spectroscopy of Milky Way UFCSs shows that systems in this regime can include both low-mass SCs and possible dwarf galaxies, demonstrating that r$_{\rm h}$ and M$_{\rm V}$ alone are insufficient for an unambiguous classification
\citep{Cerny2026}. Although all four of our candidates are more luminous than the formal UFCS luminosity range, DSC~3 extends the M82 SC population toward this regime. Their positions in the r$_{\rm h}$--M$_{\rm V}$ plane favor an SC interpretation. We therefore classify the four systems as DSCs. We also note that measurements of their radial velocities and spectroscopic metallicities would provide stronger constraints on their nature and help elucidate their origin within the M82 cluster population. 

The predominantly old stellar populations of the DSCs also distinguish them from the young and intermediate-age SC populations associated with more recent episodes of star formation in M82. The two spectroscopically confirmed old GCs identified by \citet{Saito2005} were interpreted as belonging to an older SC population, distinct from the young SCs that may have formed during the interaction with M81. In this context, the DSCs identified here are more naturally associated with the older M82 SC population than with its recent starburst activity. 

What remains less clear is how the DSCs acquired their extended structures. They may have formed with relatively large sizes, or initially more compact SCs may have expanded over several Gyr through stellar mass loss and internal dynamical evolution \citep{Gieles2010,Madrid2012,Gieles2021}. Their locations within the halo may also be important for their subsequent evolution, since the weaker tidal field away from the main body of M82 can favor the long-term survival of low-density SCs that would be more susceptible to disruption in the inner galaxy \citep{Liu2016}. An extreme example is the EC associated with the Eridanus~II UFD, also shown in Figure~\ref{fig:size_lum} using a brown cross symbol, which has M$_{\rm V}\simeq-3.5$ and r$_{\rm h}\simeq13$~pc \citep{Crnojevic2016, Simon2021}. Its survival in such a low-mass host illustrates that extended, low-luminosity SCs can persist in weak tidal environments. The present sizes of the M82 DSCs may similarly reflect both their initial formation conditions and their subsequent dynamical evolution.

Finally, the extension of the M82 SC population to such low luminosities suggests that its faint end may remain incompletely sampled. DSC~1 lies outside the HST footprint analyzed by \citetalias{Lim2013}, while DSCs~2, 3 and 4 fall within their coverage but are faint and located in relatively crowded regions, which may have contributed to their non-detection. At the distance of M82, these low surface brightness systems are only sparsely resolved and can be difficult to distinguish from the surrounding stellar field. The depth and spatial resolution of JWST allow their diffuse integrated light and resolved stars to be detected simultaneously, while also enabling better discrimination between SCs and contaminant background galaxies. Our discovery of four such systems therefore suggests that faint DSCs may remain underrepresented in current SC censuses, motivating systematic searches in nearby galaxies.

\section{Summary and Conclusions}
\label{sec:conclusions}

We report the discovery of four faint DSCs in the halo of M82 using deep archival JWST/NIRCam imaging. The systems were identified through visual inspection of the F115W and F200W images as diffuse, semi-resolved stellar systems. All four lie within the halo region of M82. Our main results are summarized as follows:

\begin{enumerate}

\item The CMDs of the four DSCs are consistent with predominantly old stellar populations, with ages of $\sim10$~Gyr. DSC~1 is consistent with a metal-poor population ($[\mathrm{M/H}]\sim-2$), whereas DSCs~2, 3 and 4 are better represented by more metal-rich isochrones ($[\mathrm{M/H}]\sim-0.8$). Given the small number of resolved stars, these age and metallicity estimates should be regarded as approximate.

\item We measure the cluster sizes and integrated magnitudes using two independent methods: a non-parametric COG analysis and PSF-convolved exponential-profile fitting. The two methods yield generally consistent results. Assuming the adopted distance to M82 (D = 3.55~Mpc), which is supported by the resolved star CMDs of the clusters, their half-light radii span r$_{\rm h}\simeq5.4$ to 10.5~pc, while the absolute magnitudes range from M$_{\rm V}\simeq-4.1$ to $-6.4$. These measurements show that all four systems are more extended than typical GCs, which have r$_{\rm h}\sim3$~pc. 

\item The newly identified DSCs extend the known M82 SC population by approximately 1 mag toward fainter M$_{\rm V}$. DSC~3, with M$_{\rm V}=-4.07^{+0.35}_{-0.36}$ and r$_{\rm h}=5.37^{+0.59}_{-0.45}$~pc, is the faintest SC candidate identified in M82 to date and approaches the size-luminosity regime occupied by Milky Way UFCSs.

\end{enumerate}

Follow-up  spectroscopy of the brighter DSCs will be particularly valuable for obtaining independent constraints on their ages and metallicities. Radial-velocity measurements would additionally provide an important test of their association with M82 and help constrain their possible origins.

The nature of this discovery based on a simple visual search suggests that faint DSCs may remain underrepresented in current SC censuses and motivates systematic searches for similar systems in nearby galaxies. The sensitivity and spatial resolution of JWST enable the detection of diffuse integrated light from these low surface brightness systems while resolving individual member stars, making faint DSCs accessible at the distance of nearby galaxies. Wide-field surveys with Euclid are already beginning to uncover faint stellar systems in nearby galaxies \citep[e.g.,][]{Larsen2025, Howell2026}, while the Nancy Grace Roman Space Telescope will greatly expand such searches by combining wide-field coverage with high spatial resolution. Together, these facilities offer the prospect of establishing how common faint DSCs are, how their properties vary across different galactic environments and eventually, how they form.

\begin{acknowledgments}

We thank A. Drlica-Wagner for an early discussion that helped motivate this work. We are also grateful to E. Peng and A. Bij for their valuable feedback on this work. DC acknowledges support from NSF grant AST-2508747. BMP acknowledges support from NSF grant AST2508745. JJ acknowledges support from the National Aeronautics and Space Administration (NASA) under Grant No. 80NSSC25K0365.

This work is based on observations made with the NASA/ESA/CSA James Webb Space Telescope. The data were obtained from the Mikulski Archive for Space Telescopes at the Space Telescope Science Institute, which is operated by the Association of Universities for Research in Astronomy, Inc., under NASA contract NAS5-03127 for JWST. These observations are associated with program GO-5145. 

\end{acknowledgments}

\facilities{JWST}

\software{ Astropy \citep{astropy2013,astropy2018,astropy2022}, 
Numpy \citep{numpy},
\texttt{DS9} \citep{DS9}}

\appendix

\section{DSC candidates in the M82-centric Coordinate System}\label{appendix1}

\begin{figure*}[!ht]

    \centering

    \includegraphics[width=\columnwidth]{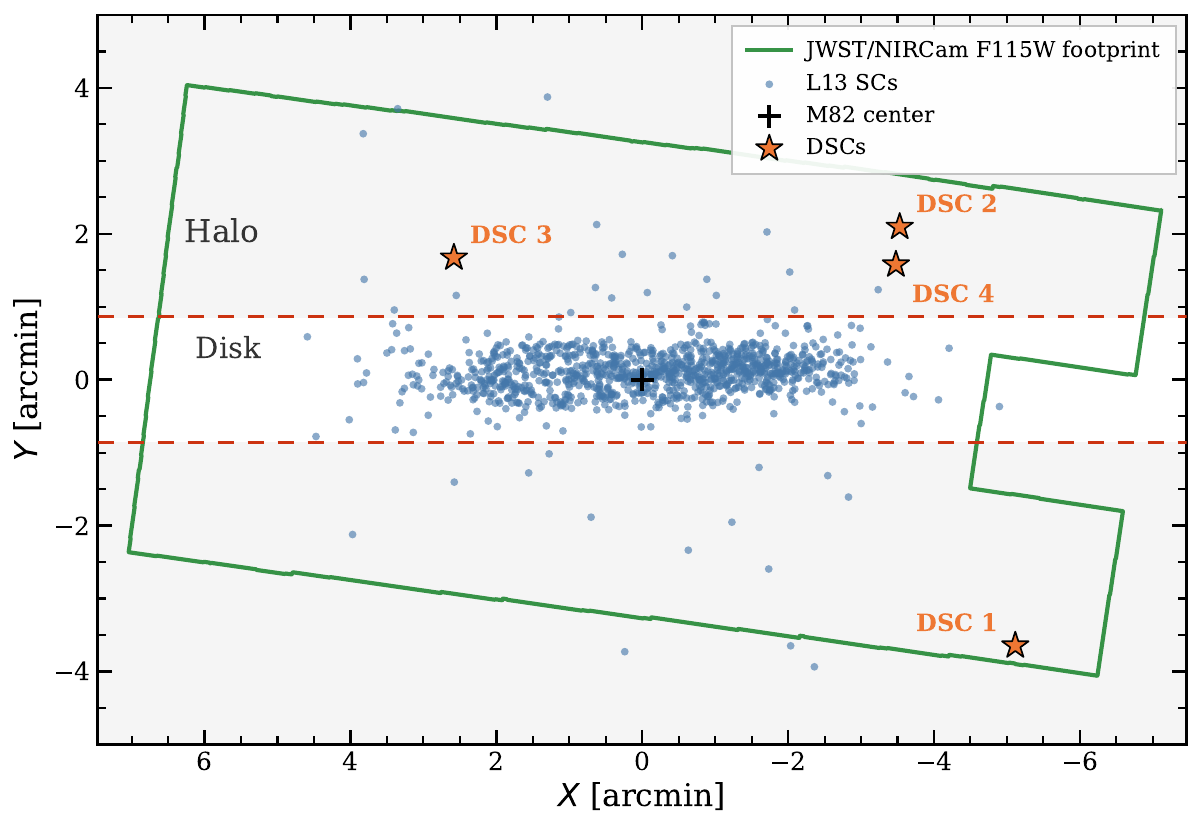} 

    \caption{Locations of the four DSCs in the M82-centric coordinate system. The $X$- and $Y$-axes are aligned with the projected major and minor axes of M82, respectively. The orange stars mark the DSCs, and the blue points show the SCs from \citetalias{Lim2013}. The black cross marks the adopted center of M82, while the green outline shows the JWST/NIRCam F115W footprint. The red dashed lines indicate the disk--halo boundary, $|Y|=3h_z=51.87\arcsec$ ($\simeq0.88$~kpc), adopted from \citetalias{Lim2013}.}

    \label{fig:xy_appendix}

\end{figure*}

Figure~\ref{fig:xy_appendix} shows the locations of the four DSCs in an M82-centric coordinate system aligned with the projected major and minor axes of the galaxy. The $X$-axis follows the major axis of M82, while the $Y$-axis represents the projected distance perpendicular to the disk. We adopt a position angle of $65^\circ$ for M82, following \citetalias{Lim2013}. The dashed lines mark the halo boundary defined by \citetalias{Lim2013}, $|Y|>3h_z=51.87\arcsec$ ($\simeq0.88$~kpc), where $h_z$ is the scale height of the thick-disk component. All four DSCs lie beyond this boundary and are therefore located within the M82 halo region according to this definition.

\section{Selection of resolved sources for the DSC CMDs}\label{appendix2}

\begin{figure*}[!ht]

    \centering

    \includegraphics[width=0.49\columnwidth]{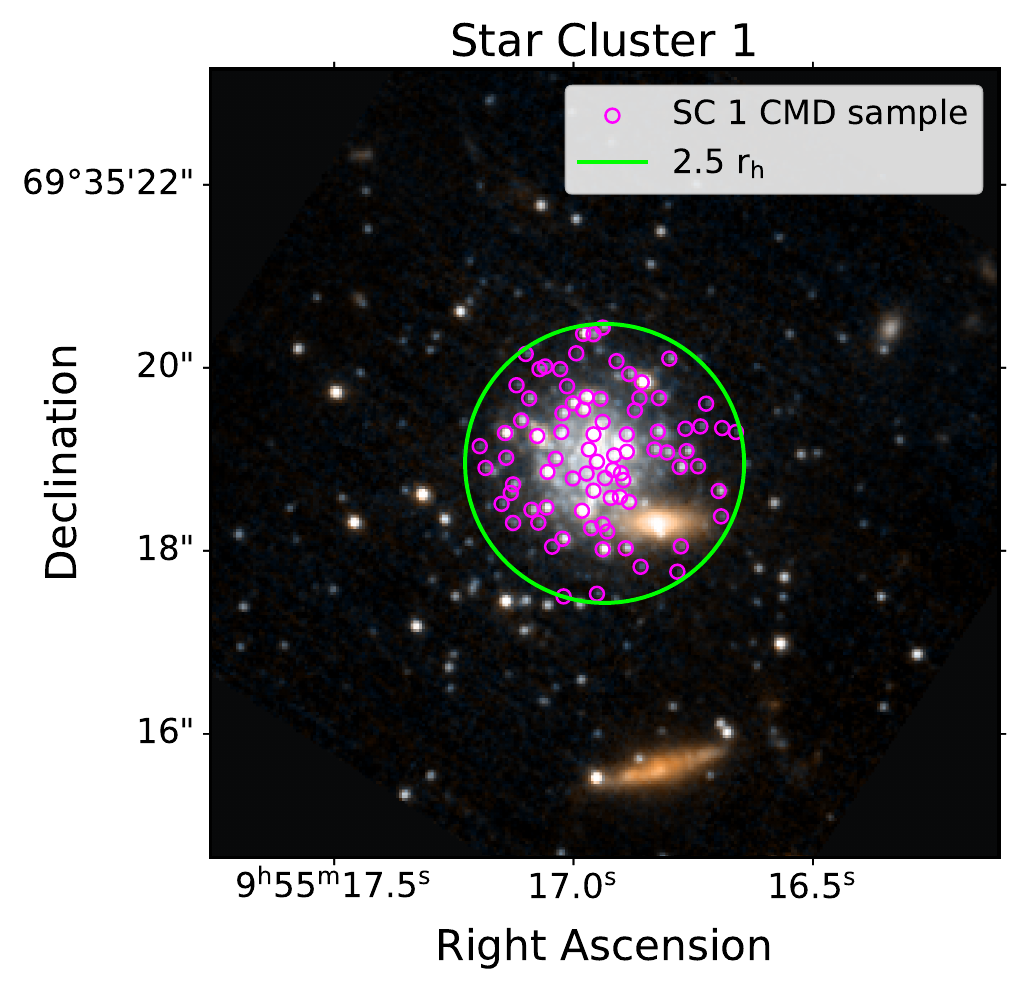} \includegraphics[width=0.49\columnwidth]{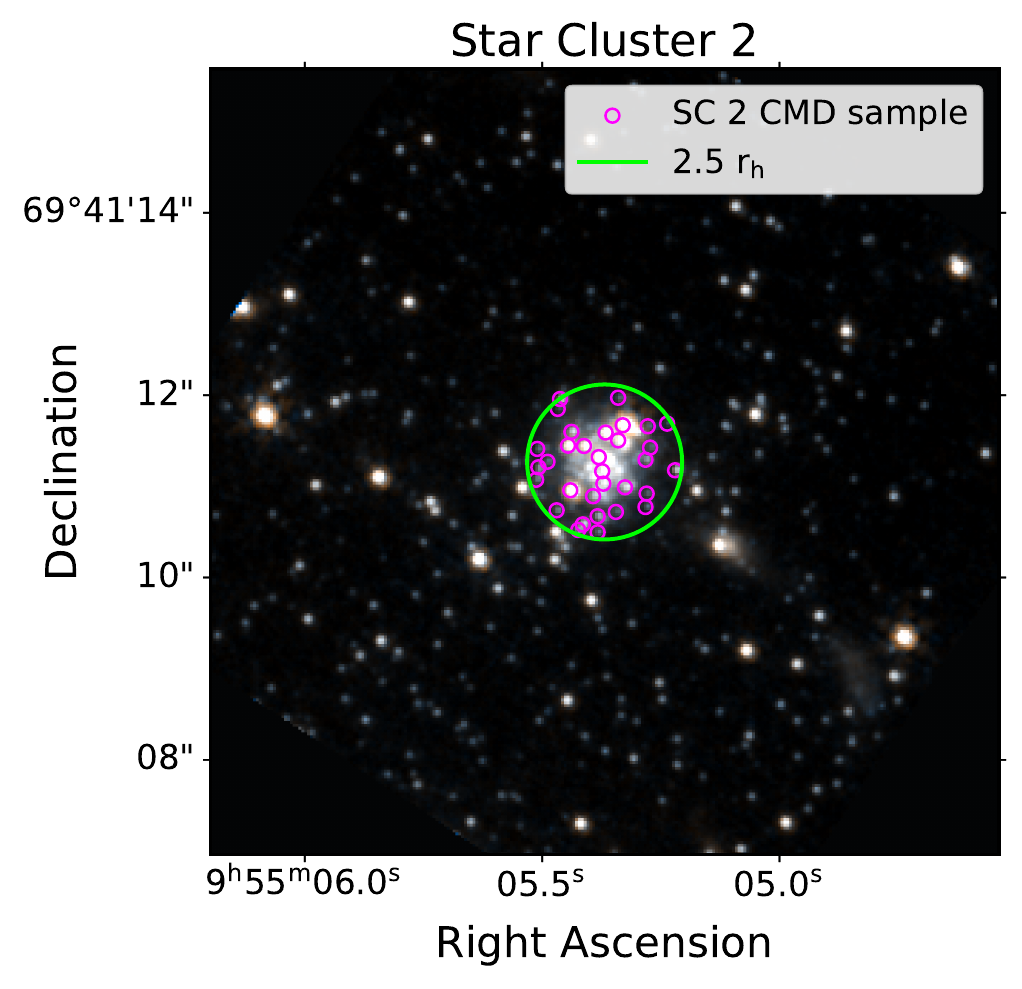} \includegraphics[width=0.49\columnwidth]{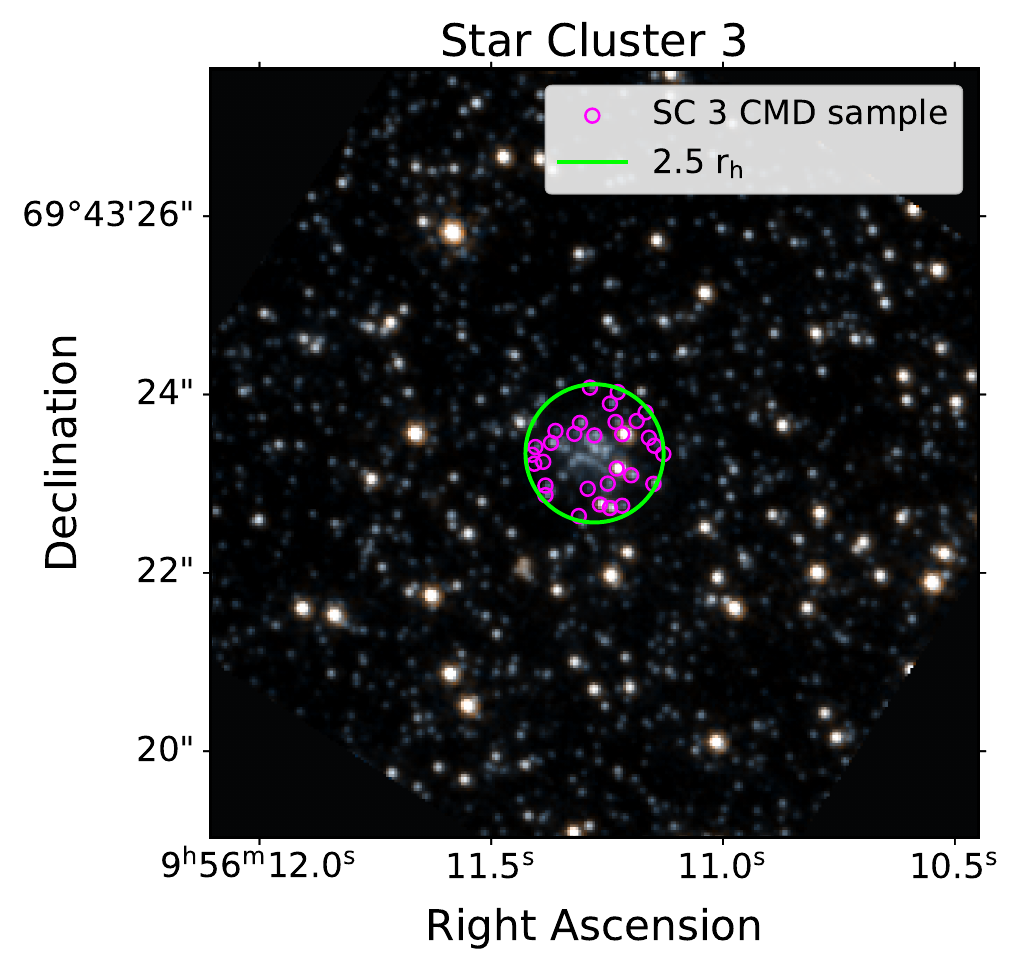} \includegraphics[width=0.49\columnwidth]{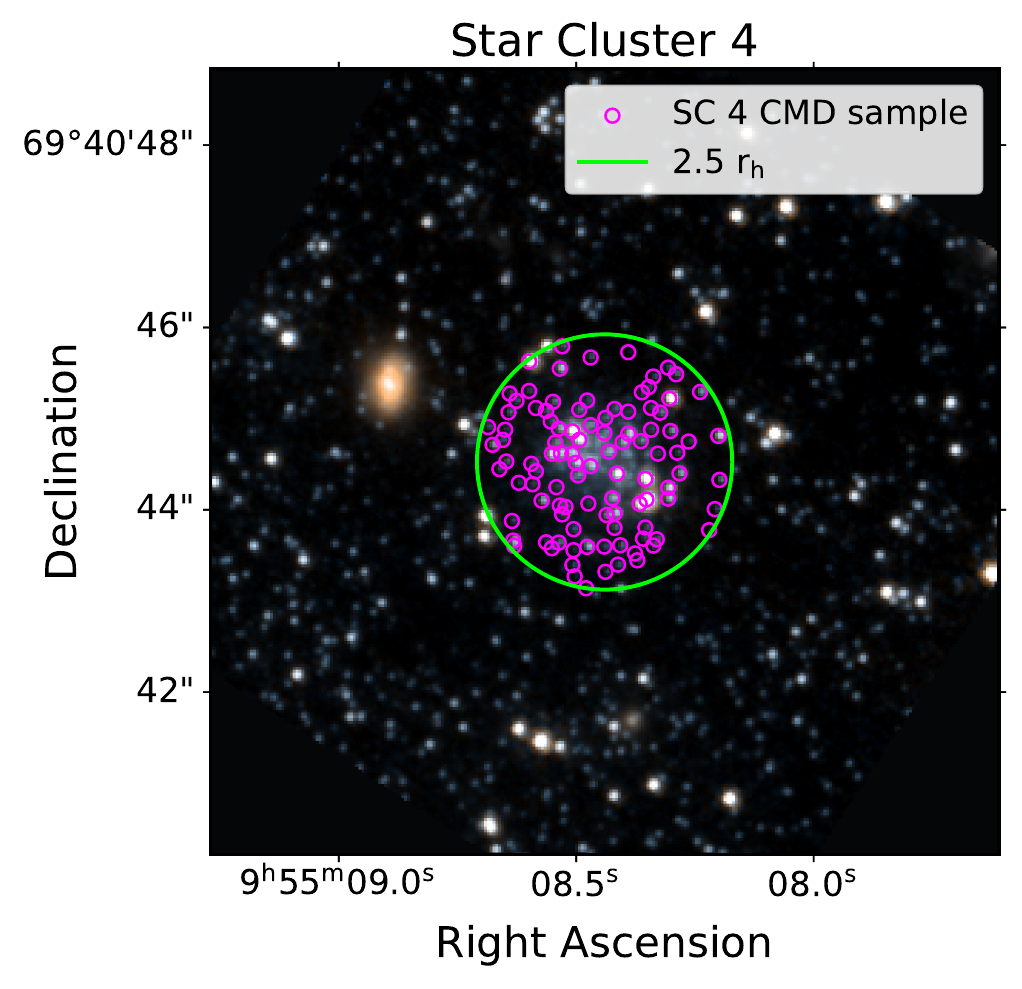} 

    \caption{Selection of resolved stars used to construct the CMDs of the four DSCs. The magenta circles mark the sources included in each CMD, overlaid on the corresponding JWST/NIRCam color image. The green circle indicates the 2.5\,r$_{\rm h}$ selection radius adopted for each system.}

    \label{fig:cmd_sel}

\end{figure*}

Figure~\ref{fig:cmd_sel} shows the resolved sources used to construct the CMDs of the four DSCs. The sources are overlaid on the corresponding JWST/NIRCam color images to illustrate their spatial distribution. For each DSC, the CMD sample is selected within 2.5\,r$_{\rm h}$ of the cluster center, as described in Section~\ref{sec:cmd}.

\bibliography{refs}{}
\bibliographystyle{aasjournalv7.1}

\end{document}